\documentclass[12pt]{article}

\usepackage{scicite}
\usepackage{graphicx}
\usepackage{times}
\usepackage{float}
\usepackage{amsmath}
\usepackage{color}

\newenvironment{sciabstract}{%
\begin{quote} \bf}
{\end{quote}}

\title{Cathodoluminescence excitation spectroscopy: nanoscale imaging of excitation pathways}

\author
{Nadezda Varkentina,$^{1\ast}$ Yves Auad,$^{1\ast}$ Steffi Y. Woo,$^{1}$  Alberto Zobelli,$^{1}$\\
Laura Bocher,$^{1}$ Jean-Denis Blazit,$^{1}$ Xiaoyan Li,$^{1}$ Marcel Tenc\'e,$^{1}$\\
Kenji Watanabe,$^{2}$ Takashi Taniguchi,$^{3}$ Odile St\'ephan,$^{1}$ Mathieu Kociak,$^{1\dagger}$\\
Luiz~H.~G.~Tizei,$^{1\dagger}$\\
\normalsize{$^{1}$Universit\'e Paris-Saclay, CNRS, Laboratoire de Physique des Solides,} \\
\normalsize{91405, Orsay, France}\\
\normalsize{$^{2}$Research Center for Functional Materials, National Institute for Materials Science,} \\
\normalsize{1-1 Namiki, Tsukuba 305-0044, Japan}\\
\normalsize{$^{3}$International Center for Materials Nanoarchitectonics,}\\
\normalsize{National Institute for Materials Science, 1-1 Namiki, Tsukuba 305-0044, Japan}\\
\normalsize{$^\ast$These authors contributed equally}\\
\normalsize{$^\dagger$To whom correspondence should be addressed;}\\
\normalsize{E-mail:  luiz.galvao-tizei@universite-paris-saclay.fr/}\\
\normalsize{mathieu.kociak@universite-paris-saclay.fr}
}

\date{}

\begin{document} 


\baselineskip24pt


\maketitle


\begin{sciabstract}

Following the lifespan of optical excitations from their creation to decay into photons is crucial in understanding materials optical properties. Macroscopically, techniques such as the photoluminescence excitation spectroscopy provide unique information on the photophysics of materials with applications as diverse as quantum optics or photovoltaics. Materials excitation and emission pathways are affected by nanometer scale variations directly impacting devices performances. However, they cannot be directly accessed, despite techniques, such as optical spectroscopies with free electrons, having the relevant spatial, spectral or time resolution. Here, we explore optical excitation creation and decay in two representative optical devices: plasmonic nanoparticles and luminescent 2D layers. The analysis of the energy lost by an exciting electron that is coincident in time with a visible-UV photon unveils the decay pathways from excitation towards light emission. This is demonstrated for phase-locked interactions, such as in localized surface plasmons, and non-phase-locked ones, such as the light emission by individual point defects. This newly developed cathodoluminescence excitation spectroscopy images energy transfer pathways at the nanometer scale. It widens the toolset available to explore nanoscale materials.

\end{sciabstract}

Light emission spectroscopies reveal materials' optical excitations. The knowledge of the mechanisms leading from light absorption to emission, i.e, the absorption and decay pathways, is paramount for understanding these excitations' physics and their applications. Photoluminescence excitation spectroscopy (PLE) is especially suitable for this purpose. In this spectroscopy, the emission intensity is measured as a function of excitation energy. Quantitatively, it directly maps a system's relative quantum efficiency as a function of excitation energy. Qualitatively, it permits to access the competition between different relaxation pathways from the selected absorption states towards emission in the selected energy window.

PLE has proven to be invaluable as it provides unparalleled information on the optical properties of materials. Examples include identification of excited exciton states and quantitative measurement of their binding energy in 2D materials \cite{Hill2015, Robert2018}, determination of the energy transfer quantum efficiency in carbon nanotube/porphyrin compounds \cite{Roquelet2010}, exhaustive characterisation of the photophysics of single photon emitters in nanodiamonds \cite{Beha2012} and defects in h-BN \cite{Museur2008}, and deep insight into the relaxation pathways in GaAs quantum dots \cite{Brunner1992}. The correlative nature of PLE makes it  extremely sensitive compared to other absorption techniques.  Despite all these advantages and their impact in all fields relying on optical material characterization, from quantum optics to photovoltaics, light diffraction imposes a limit onto spatial resolution for PLE, to within few hundreds of nanometer at best. This severely hinders its application, as the efficiency of the excitation and decay pathways vary drastically at scales much smaller than the wavelength of light \cite{Narukawa1997}. 

Free electron based microscopies may potentially solve this issue, benefiting from  sub-optical wavelength spatial resolution due to the small wavelength of electrons (3.6 pm for 100 keV electrons), and of broadband excitation extending from the infrared to the hard X-ray range \cite{E96}.

To start with, cathodoluminescence (CL) is an emission spectroscopy \cite{HY1986, Zagonel2011, Polman2019} that measures the light emission spectrum under free-electron excitation.  The last 20 years have witnessed an impressive success of this technique for nanosciences \cite{Kociak2017, Polman2019}, because it can be seen as a nanoscale equivalent of off-resonance PL for semiconductors \cite{Mahfoud2013}, and of scattering spectroscopy for plasmonic and optical excitations \cite{paper035,paper149, paper251}. Nevertheless, as the electron excitation is not monochromatic, a cathodoluminescence excitation spectroscopy (CLE) could not be developed solely by mimicking the principles behind PLE.   

Attempts to circumvent this problem includes the introduction of a novel light intensity auto-correlation method in CL. It indicated that at least for some excitation pathways the bulk plasmon creation and decay into multiple electron-hole pairs have to play a role \cite{MTC15}, as previously proposed \cite{Rothwarf1973}. This technique allows the nanoscale mapping of the energy-integrated relative quantum efficiency of semiconductor nanowires \cite{Meuret2017, Meuret2018}, without resolving the absorption energy at the origin of the luminescence, therefore failing to resolve the exact  physical origin of the absorption and decay pathways. Nevertheless, the plethora of other possible pathways to emission has not been investigated nor considered.

To solve this problem, this absorption information can in principle be retrieved with a companion relativistic electron spectroscopy, the electron energy loss spectroscopy (EELS). It can be described as a nanometer scale counterpart of absorption (or more precisely extinction) spectroscopy \cite{paper251}. It has been used in combination with CL \cite{paper251, Bonnet2021, Auad2022} to gain insights into the physics behind light emission upon electron scattering. The attribution of a given emission event (CL) to the creation of an excitation at a given energy is an information bear by individual electrons in an EELS spectrum. Unfortunately, it is lost with the current time-averaged technologies, making impossible the investigation of excitation to emission pathways at the nanometer. CLE was therefore still to be invented.

Here, we demonstrate CLE with nanometer scale spatial resolution over a broad energy range (from the visible to the soft X-ray, 2 to 620 eV) in a scanning transmission electron microscope (STEM). Our approach relies on a new coincidence scheme between inelastic electron scattering and photon emission events. If the temporal information of both these events is known, correlation can be performed to unveil the probability of each energy transfer pathways. CLE spectra are constructed with EELS events that are time-correlated with a photon emission, while energy relative quantum efficiency spectra are given by the ratio of CLE and total EELS spectra. As a proof-of-principle of CLE, we focused on representatives of the two main families of optically relevant materials, plasmonic nanoparticles for the photonic materials and defects in semiconducting materials for the luminescent ones. Studying CLE on Au nanospheres embedded in SiO$_2$, two light emission pathways are identified: surface plasmons for the Au (SP) and transition radiation (TR) for the SiO$_2$ and Au. The direct energy and time correlation between absorption and emission for these excitations, known to be phase-locked with the exciting electron, is a confirmation of the relevance of CLE. With CLE on \textit{h}-BN flakes, TR was also detected. This signal is usually in the background of EELS spectra, evidencing the extremely high sensitivity of the technique, with a typical 2 orders of magnitude improvement. CLE was more importantly used to explore the decay pathways leading to the excitation and emission of the 4.1 eV defect in \textit{h}-BN. All excitations, from the near-band edge (NBE) to the core-losses, including the bulk plasmon, are demonstrated to participate to photon emission. The bulk plasmon is experimentally confirmed as the main absorption pathway. Nevertheless, the relative quantum efficiency first peaks at the NBE energy and is followed by a linear increase up to the maximum energy in the soft X-ray energy range (620 eV), which had not yet been observed. The NBE pathway is unexpectedly the most efficient excitation channel for defect light emission, up to absorption energies of 15 eV.  Finally, spatially resolved CLE in \textit{h}--BN reveals the spatial variation of the excitation and decay pathways with a 125 nm spatial resolution. STEM-CLE, on that account, has proven to be a nanometer-scale counterpart of PLE.

\begin{figure}
\begin{center}
  \includegraphics[width=6 in]{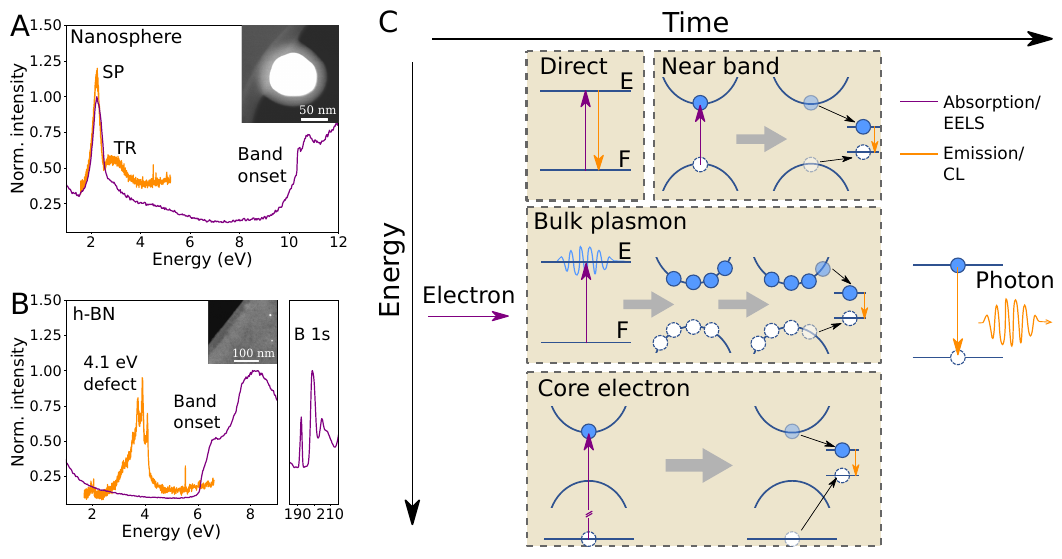}\\
  \caption{ \textbf{Photon emission pathways upon electron scattering}:  \textbf{(A-B)} Time-averaged CL (orange) and EELS (purple) spectra of an Au/SiO$_2$ nanosphere and a thin \textit{h}--BN flake show different absorption and emission features, described in the text. From these correlative time-averaged spectra, one cannot identify which absorption transitions lead to light emission. The small intensity emission at $\approx$2 eV in the h-BN CL spectrum is a replica of the 4.1 eV defect emission due to the diffraction grating. The insets show images of the nanosphere and the h-BN edge. CL and EELS spectra have been normalized and shifted vertically for clarity. \textbf{(C)} A relativistic inelastic electron scattering event in a solid can generate different excitations (vertical purple arrows): direct optical transition, NBE transition, bulk plasmon excitation, and core-level transitions. Excitations not involving single particles (excitons, bulk and surface plasmons, etc.) are represented between a fundamental (F) and excited (E) state. These can relax through different pathways, leading to the excitation of a final optically bright energy level and to photon emission (vertical orange arrows).}
  \label{Exc-Dec_Pathways}
\end{center}
\end{figure}

In the following, we concentrate on Au nanosphere embedded in SiO$_2$, mainly showing SP resonances at around  2.2 eV both in absorption and emission, and \textit{h}-BN flakes, with a dominating NBE and plasmonic absorption and strong 4.1 eV defect emission \cite{Taniguchi2007, Bourrellier2016}, as clearly seen on the absorption (EELS) and emission (CL) spectra in Fig. \ref{Exc-Dec_Pathways}A and B. Nevertheless, due to their time-averaged nature, these spectra alone cannot  reveal the excitation to emission pathways shown schematically in Fig. \ref{Exc-Dec_Pathways}C.


Indeed, electron scattering in matter leads to light emission through different processes, extending in a wide energy (from meV to keV) and time (from fs to $\mu$s) ranges. In Fig. \ref{Exc-Dec_Pathways}C, optical transitions are represented by vertical arrows and relaxation pathways by black arrows, with qualitative temporal axis from left to right.

In order to understand CLE, it is necessary to know how EELS and CL spectra relate to the processes described in Fig. \ref{Exc-Dec_Pathways}C. Every inelastically scattered electron must undergo an initial excitation (purple arrow in Fig. \ref{Exc-Dec_Pathways}C) that can be measured with EELS. This encompasses TR, NBE excitations, core-level excitations \cite{E96}, bulk \cite{R1980} and surface \cite{B1982} plasmon  excitations, phonon excitations \cite{IM1982,Krivanek2014} and exciton excitations \cite{Bonnet2021}. TR occurs when a relativistic electron crosses a dielectric discontinuity and is often missed in the presence of other excitations in the same energy range, due to its small oscillator strength (Fig. \ref{Exc-Dec_Pathways}A-B). The NBE of semiconductors is easily detected in EELS (Fig. \ref{Exc-Dec_Pathways}A-B for SiO$_2$ and \textit{h}-BN), especially with modern electron monochromator technologies \cite{Krivanek2014}. Core-electron spectroscopy is widely used for chemical mapping and allotrope identification \cite{E96} down to the atomic scale \cite{Muller2008}.

After having been created through the above-detailed absorption process revealed by EELS, these distinct excitations over a wide range of energies can lead to photon emission, detected with CL in the IR-UV range, through different relaxation pathways, some of which are still not understood. TR and SP are typical of photonic materials, characterized by a phase-locked emitted photon relative to the exciting electrons \cite{paper149}. As a consequence extinction (EELS) and emission (CL) spectra are similar, with only slight shifts expected\cite{paper251}, and therefore we expect a CLE spectrum to closely resemble a CL spectrum. In luminescent materials, absorption and decay pathways are expected to be more complex upon electron excitation. As depicted in Fig. \ref{Exc-Dec_Pathways}C, NBE, bulk plasmons, core-level excitations, or direct excitations can lead to the emission of light, and we expect the CLE to be quite different from the EELS. A microscopic description of the weight of each energy transfer processes is still not available.

\begin{figure}
\begin{center}
  \includegraphics[width=4.6 in]{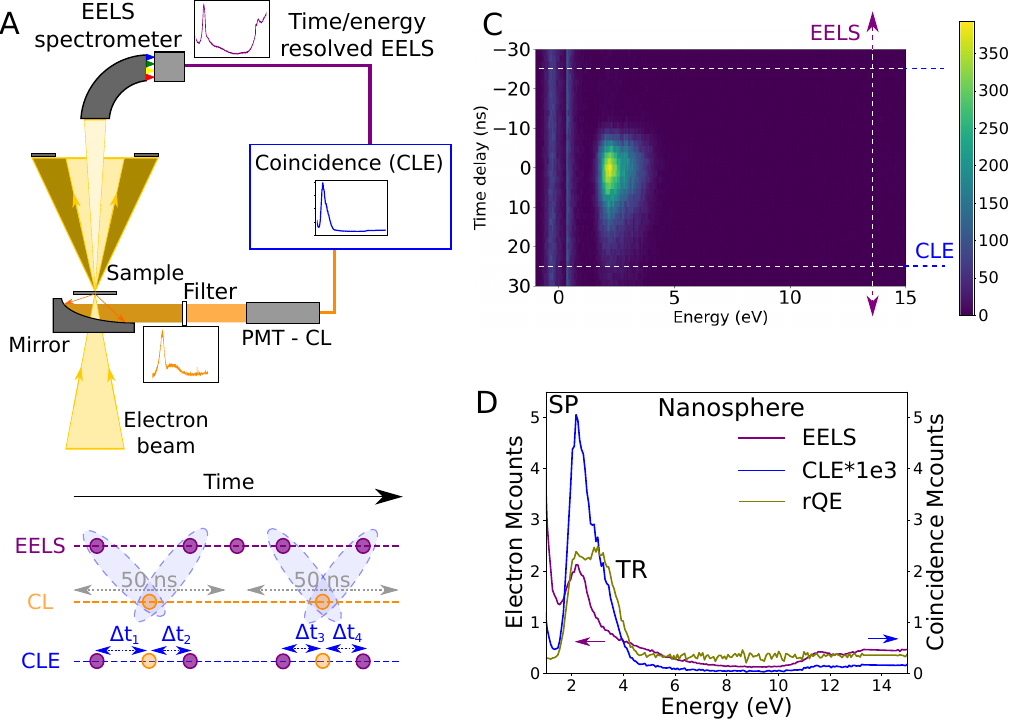}\\

  \caption{\textbf{CLE in a STEM}  \textbf{(A)} Sketch of the experimental setup: A 60-100 keV electron beam is focused in a nanometer spot that can be scanned along the surface of a sample. Time-resolved CL events (orange) are collected through a parabolic mirror and detected, after passing through a filter, with a PMT, and time-resolved EELS events are measured by a TimpePix3 detector after an EELS spectrometer. These are stored in an ordered list, used to produce coincidence spectra (blue). \textbf{(B)} A search algorithm is used to find electrons which are within $\pm 25$ ns of a detected photon, from which a 2D histogram of time delay versus electron energy loss is reconstructed. \textbf{(C)} This 2D histogram shows the temporal evolution of the loss spectrum a as function of delay to a detected photon.  \textbf{(D)} With this information, a total EELS (all detected electrons), the CLE ($\pm 5$ ns from $\Delta t = 0$ ns) and the relative quantum efficiency for a Au/SiO$_2$ nanosphere was calculated. For the nanosphere the SP and TR decay channels are efficient photon emission pathways. TR is less visible in the average EELS spectrum, it is exacerbated in the CLE.}
  \label{CLE_Plasmons}
\end{center}
\end{figure}

 An emission (CL) event is necessarily preceded by an absorption or extinction event at a given energy (EELS). This relation is temporal in nature, and is lost in commonly time-averaged EELS spectra where all potential EELS events corresponding to the same emission are summed. It is however stored in the probability of each electron scattering event and photon emission. This can be retrieved by generating coincidence histograms of electron energy loss and photon emission events (described in what follows, Fig. \ref{CLE_Plasmons}A-B, and in the SI).  Coincidence electron spectroscopy and microscopy have been performed in the past, for example coincidence of EELS with secondary electron or X-ray emission \cite{Kruit1984,Jannis2019,Jannis2021}).  EELS-CL coincidence have been performed for discrete selected EELS energy ranges \cite{Ahn1984, Graham1986}, but the relative quantum efficiency as a function of energy and its spatial dependence has not been measured. 

To achieve CLE, a temporal resolution below the time interval between events, given by the electron current (typically 1 electron every 25 ns for 10 pA), is required for all energy loss events of interest. With this in mind, we implemented an EELS-CL set-up in a STEM, displayed in Fig.  \ref{CLE_Plasmons}A. In Fig. \ref{CLE_Plasmons}, we illustrate the principle of CLE on the simplest case of the plasmonic particles. For EELS, a Timepix3 detector was used \cite{Auad2022b}. The detector provides sub 10 ns time resolution over arbitrary energy ranges determined by the resolution power of the electron spectrometer and the Timepix3 pixel size. Additionally, the particular detector used (CheeTah, from Amsterdam Scientific Instruments) has two time-to-digital converters (TDCs), allowing to append timestamps from external signals into the original electron data flow. Photon emission events were detected with a photomultiplier tube (PMT) working in the 2.0 to 5.0 eV energy range. The PMT output is directly connected to one of the Timepix3 TDC lines. Electrons and photons arrival times were stored in a list, along with the electron energy loss. The response time of the detection scheme is $\approx$5 to 25 ns. A search algorithm (\textbf{Methods} in the SM and code available at Zenodo \cite{Auad2022_6346261}) is used to find electrons which are within $\pm 25$ ns of a detected photon, from which a 2D histogram of time delay versus electron energy loss is reconstructed, Fig. \ref{CLE_Plasmons}C. This 2D histogram shows the temporal evolution of the loss spectrum as a function of delay to a detected photon.

From these, a 2D histogram of electron energy loss events as a function of time delay to a photon emission is reconstructed (Fig. \ref{CLE_Plasmons}C). Because of the typical lifetimes of the CL events (SP and TR sub-ps and defect emission sub-ns), the CLE spectrum is extracted from the shortest time delays given the time response of the experiment ($\pm 5$ ns), within which coincidence above the long delay limit was observed. For longer lifetimes, larger time integration should be considered. The CLE spectrum resembles an EELS spectrum, but weighted by the photon emission probability (Fig. \ref{CLE_Plasmons}D). Finally, the ratio of the CLE and the non-coincidence EELS therefore provides the relative quantum efficiency (QE) of different absorption processes (Fig. \ref{CLE_Plasmons}D). It highlights differences between competitive radiative and non-radiative pathways.

For the nanospheres (Fig. \ref{CLE_Plasmons}D and Fig. \ref{SameAxisNanoSFig2}-\ref{CLE_nanosphere_aloof_impact}) with the electron beam incident on the SiO$_2$ shell, photon emission is due to the Au nanosphere SP decay (2.0-2.4 eV) and SiO$_2$ TR (2.6-4.0 eV), while higher energy losses do not contribute to light emission in the emission detection range.  This is a reassuring observation, as photonic modes such as plasmons or TR are created in phase with the electron field and can only be created by loss events with energies in the same range as the emission ones \cite{paper149}. In the same line, the relative QE is featureless above the SP and TR energies. This is expected as for a phase-locked excitation, we do indeed expect that all the light emitted at a given frequency to have been triggered by an extinction event at the same energy - no energy is transfer from higher frequencies. A similar observation is reported by Feist et al. on micrometric photonic structures using an equivalent EELS-CL coincidence experiment \cite{Feist2022}.  We note that in general, some spurious coincidences (SM section \ref{Spurious}) are detected, but this cannot be avoided: part of them stem from detector noise (PMT photocathode, ambient light leakage) or from the poissonian nature of the electron source used (this could be improved with a pulsed electron source or a better detector temporal point spread function). Also, the SP and the TR peaks observed are modulated by the PMT response to photons are a function of wavelength: coincidence events outside the PMT response range are missed.

\begin{figure}
\begin{center}
  \includegraphics[width=4.6 in]{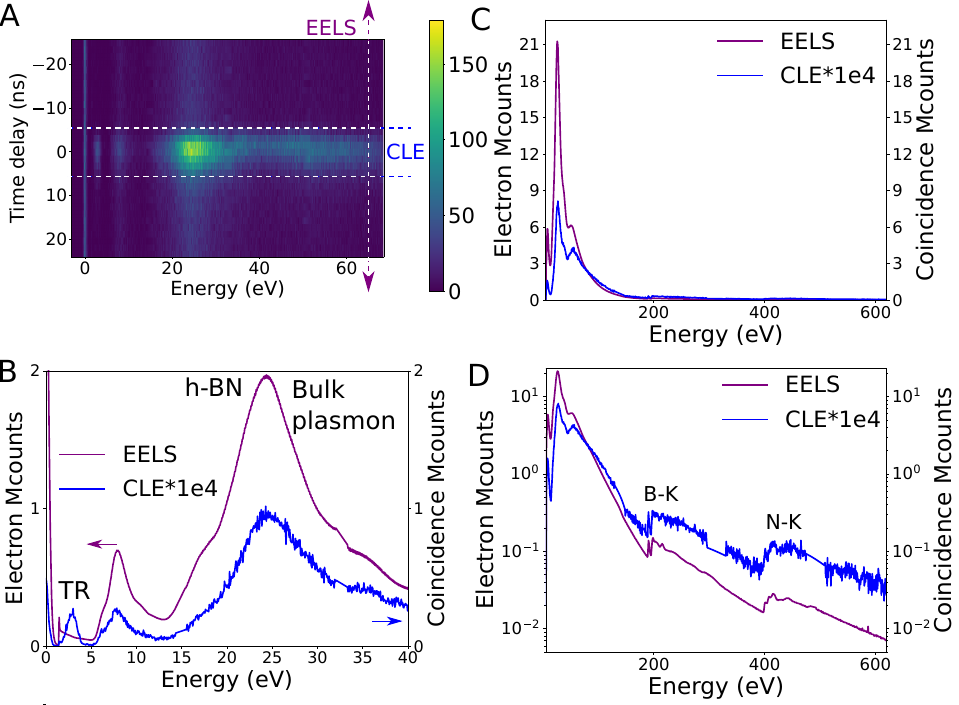}\\
  \caption{ \textbf{CLE of h-BN}  \textbf{(A)} 2D coincidence histogram for a thin \textit{h}-BN flake \textbf{(B)} EELS and CLE spectra of an \textit{h}--BN thin flake. \textbf{C-D} EELS and CL of a different \textit{h}--BN flake up to core losses in linear and logarithmic scales, respectively. In these spectra TR, NBE, and all energies above it contribute to photon emission, even up to 600 eV. Detector junctions appear at around 15 and 33 eV (panel A) and around 150, 320 and 500 eV and are interpolated in spectra (panel B).
  }
  \label{CLE_hBN}
\end{center}
\end{figure}

As the observation of these coincidence events and decay channels in the simplest case of phase-locked excitations  validate our methodology, we turn to the more involved case of semiconductors emission. In a thin \textit{h}-BN flake ($<$50 nm), the CLE spectra show contributions from TR, NBE, bulk plasmons and higher energies towards light emission (4.1 eV defect and TR, included in our detection range). As discussed, the contribution from TR is usually missed in EELS spectra because of its small cross section. As a matter of fact, they are invisible in the EELS and CL spectra of Fig. \ref{Exc-Dec_Pathways}B.  In the CLE spectra (Fig. \ref{CLE_hBN}A-B) their contribution (at 1e-5 event counts compared to the regular EELS counts, a typical 2 orders of magnitude better sensitivity than previously demonstrated) is revealed, as a signature of the high sensitivity of CLE, much along the lines of PLE. The emission of the 4.1 eV defect is peaked between 3.65 and 4.1 eV, while that of TR is much broader. The use of a broadband filter (3.65 to 4.1 eV) filters out part of the TR contribution (Fig. \ref{CLE_hBNFiltered}).

The contribution from the defect emission can be seen in the NBE, bulk plasmon (Fig. \ref{CLE_hBN}B) and core-losses up to the maximum detected energy (Fig. \ref{CLE_hBN} and \ref{CLE_core-loss_hBN}). From the CLE, we prove experimentally the common assumption that the bulk plasmon (24.4 eV) is indeed the main source of electron-hole pairs which, after relaxation, leads to emission of the 4.1 eV defect. Nevertheless, the NBE absorption is demonstrated to be a non-negligible source of emission, and core-losses to be also a possible excitation path. 

\begin{figure}
\begin{center}
  \includegraphics[width=4.6 in]{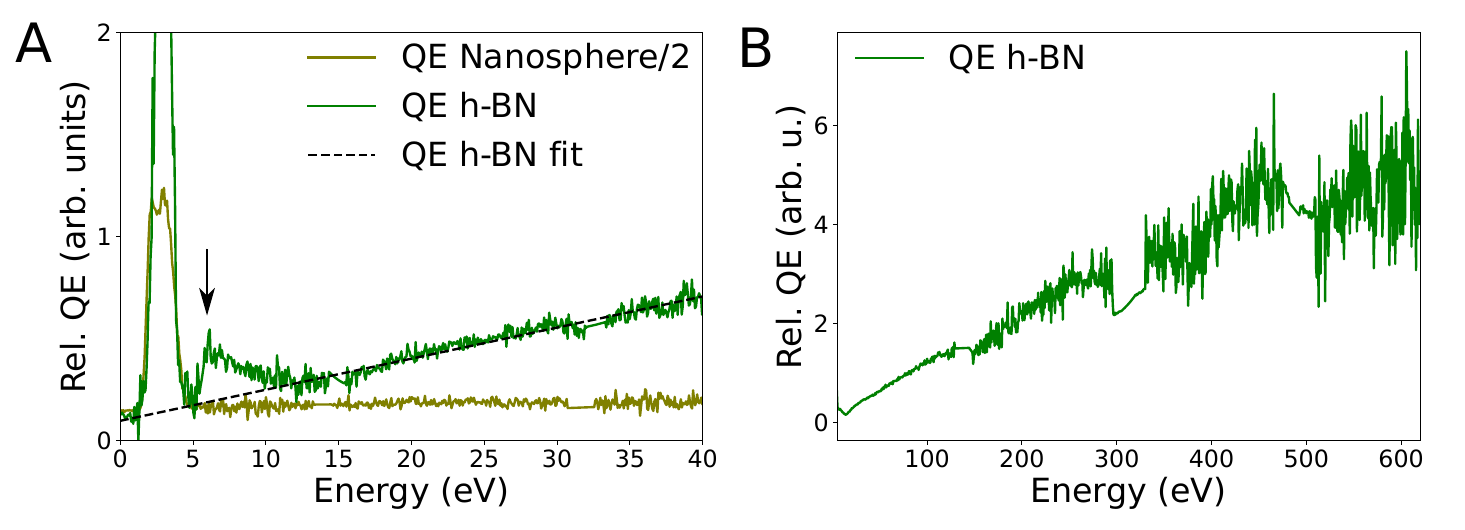}\\

  \caption{\textbf{Relative quantum efficiency}  \textbf{(A)} Relative QE of a nanosphere (Fig. \ref{CLE_Plasmons}) and a thin h-BN flake (Fig. \ref{CLE_hBN}). NBE losses are more efficient pathways for light emission than energies below the bulk plasmon. Above 15 eV, the relative QE increases linearly. The nanosphere relative QE was divided by 2 for clarity, which is shown in detail in Fig. \ref{SameAxisNanoSFig2}.   \textbf{(B)} Thin h-BN relative QE up to core losses, showing the B and N K-edges. Detector junctions appear at around 15 and 33 eV (panel A) and around 150, 320 and 500 eV and are interpolated in spectra (panel B).}
  \label{CLE_rQE}
\end{center}
\end{figure}

The investigation of energy-resolved relative quantum efficiency (Fig. \ref{CLE_rQE}) permits to better understand the physics of energy transfer from absorption to emission in semiconductors. First, we see that contrary to phase-locked excitations, for which the high energy relative quantum efficiency is completely zero, that related to the 4.1 eV defect has a non-zero and non-monotonic behaviour. Second, NBE is strikingly a more efficient excitation channel for the emission of the 4.1 eV defect than other excitations up to absorption energies of 15 eV. Above this energy, the efficiency for photon emission increases linearly, up to the maximum energy we have measured (620 eV), i.e. over an energy range much larger than achievable with PLE. The extrapolated linear trend at low energy crosses zero at the band gap energy. This can be tentatively explained as follows. Each excitation at energy loss $E$ can lead to the generation of at most $N$ electron-hole pairs, and then at most N photons, where $N= E/E_g$ and $E_g$ is the band gap energy. Below the bandgap energy the number of electron-hole pairs generated is zero. The optical bandgap of h-BN measured using EELS is around 6.0 eV \cite{Arenal2005,Liu2016}. Assuming that the last step to 4.1 eV defect emission are NBE electron-hole pairs, the linear trend is deduced. With this, the peak in the relative QE at the NBE energy is reminiscent of an unforeseen resonant effect that will require further theoretical investigation. Direct resonant excitation of defect states by fast electrons has yet to be observed.

Now that the principle of the CLE is established, we turn to the possibility of mapping the different pathways directly in real space. Indeed, the proposed spectroscopy scheme allows for coincidence mapping, of which more details are reported elsewhere \cite{Auad2022b}. The 4.1 eV emission in \textit{h}--BN (Fig. \ref{Exc-Dec_Pathways}C) is known to arise from single point defects \cite{Bourrellier2016}. For each single defect the CL excitation area forms an intensity spot $\approx$ 80x80 nm$^2$ wide. We performed CLE mapping by rastering a nanometer-sized beam on the sample, and collecting a full CLE spectrum corresponding to emission in the 3.65-4.13 eV range at each pixel of the scan. From this, CLE maps can be created by filtering over different absorption (EELS) ranges.

\begin{figure}
\begin{center}
  \includegraphics[width=5 in]{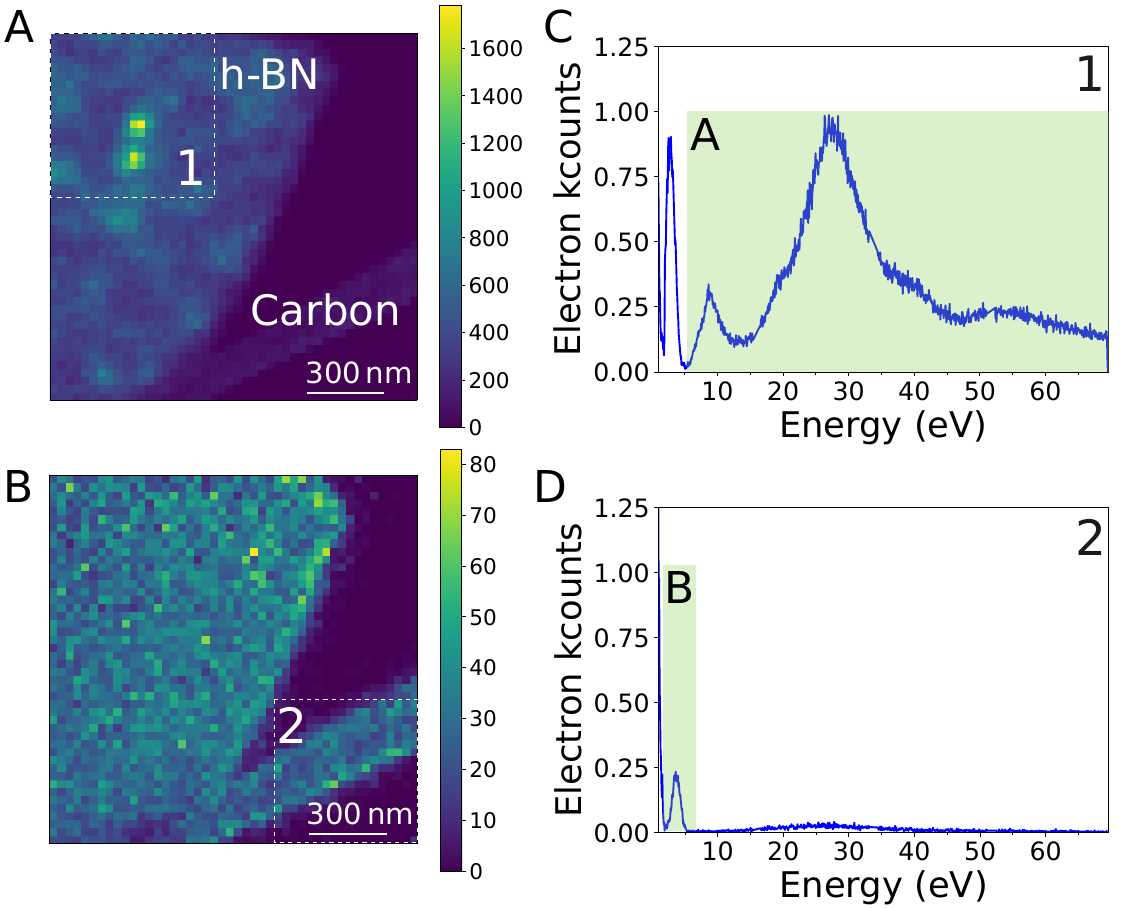}\\
  \caption{\textbf{Spatially resolved CLE maps in \textit{h}--BN}: \textbf{(A)} CLE energy filtered map above 6.5 eV, the NBE energy, showing multiple localized absorption maxima, that lead to emission of the 4.1 eV defect.  \textbf{(B)} CLE energy filtered map between 2 and 5 eV, showing where TR occurs. Both the \textit{h}--BN thin flake and the amorphous carbon support (lower right, which is the support for \textit{h}--BN in the TEM sample) show absorption leading to photon emission. \textbf{(C-D)} CLE spectra of regions marked 1 and 2 in panels A-B, with marked integrated ranges for maps A and B, respectively.}
  \label{Spim_hBN}
\end{center}
\end{figure}

These time-resolved maps permit disentangling the different decay pathways in space and energy, with a 32 nm spatial sampling. The two bright features in the image are separated by 125 nm. The  CLE map filtered above 6.5 eV energy loss shows two sharply localized intensity spots consistent with the observation of 4.1 eV  localized defects (Fig. \ref{Spim_hBN}A). On the contrary, the CLE map filtered between 2 and 5 eV (Fig. \ref{Spim_hBN}B), on the peak linked to TR,  shows that both the \textit{h}-BN flake and the thin amorphous carbon support (of the TEM grid, see Methods in the SI) exhibit coincidence events distributed in a relatively uniform manner. We note that we could not identify any specific absorption signature of the defects at their absorption energy. Coincidence measurements with better EELS spectral resolution might reveal it in the future. Also, the spatial resolution is essentially dependent on that of the CL, which is limited by the diffusion lengths in the materials. One can expect few nanometers spatial resolution in other materials, such as III-N heterostructures \cite{Kociak2017}.

In conclusion, we demonstrated spatially resolved CLE, which encompasses the main advantages of PLE (high sensitivity measurement of the relative quantum efficiency and consequent insight between multiple light emission decay pathways) with that of electron spectroscopies (wide energy range, nanometer scale spatial resolution). Numerous applications of CLE are expected for nanomaterials, ranging from the optimization of single photon sources \cite{Beha2012}, the unveiling of the role of nanometer to atomic scale features on the optical properties of transition metal dichalcogenide monolayers by mapping the excitons biding energy  \cite{Hill2015, Robert2018} to the characterization of new optical materials such as hybrid perovskites \cite{Zelewski2019}, and others yet to be discovered and understood. The spectromicroscopy scheme described requires only time-resolved electron and photon detectors, being implementable in any electron microscope. Therefore, it is applicable to any object compatible with STEM observation, should they be photonic (plasmonics systems, photonic band gap materials, waveguides ...) or luminescent (quantum wells, quantum dots, single photon emitters ...) \cite{Kociak2017, Polman2019}. The current applications of the setup in the time domain is limited by the electron detector temporal resolution. Improvements in the near future are expected with the new Timepix4 detector\cite{Campbell2016}, with fast deflectors or with the use of pulsed electron guns \cite{BFZ09, FES15, Houdellier2018}. Photon and electron energy resolved experiments in the core-level range with better temporal resolution should give further hints on the microscopic physics behind the relaxation pathways. Additionally, as the number of emitted photons per electron per energy is lost using single pixel detectors, the use of multiple PMTs or 2D arrays of detectors solves this, giving access to excitation-energy-resolved Hanbury Brown and Twiss interferometry \cite{Tizei2013} for energy-resolved retrieval of quantum statistics, energy efficiency for total photon yield and excited-energy-resolved bunching experiments \cite{MTC15}. As for PLE, such technique resolved in emission and absorption energy will allow one to assign specific energy bands to each observed transition, but now with nanoscale spatial resolution. Finally, polarization dependent EELS \cite{Guzzinati2017,Lourencco2021} and CL will give us an almost ideal nano-optics to probe excitations symmetries.



\bibliographystyle{Science}

\bibliography{CLE.bib}

\section*{Acknowledgement}

This project has been funded in part by the National Agency for Research under the program of future investment TEMPOS-CHROMATEM (reference no. ANR-10-EQPX-50), the JCJC grant SpinE (reference no. ANR-20-CE42-0020), the BONASPES project (ANR-19-CE30-0007) and the JCJC IMPULSE, (reference no. ANR-19-CE42-0001), and by the European Union’s Horizon 2020 research and innovation programme under grant agreement No. 823717 (ESTEEM3) and 101017720 (EBEAM). K.W. and T.T. acknowledge support from the Elemental Strategy Initiative conducted by the MEXT, Japan (Grant Number JPMXP0112101001) and JSPS KAKENHI (Grant Numbers 19H05790, 20H00354 and 21H05233). The authors thank ASI for extensive discussion during the Timepix3 CheeTah implementaiton for EELS and coincidence experiments. 

\section*{Authors contributions}

LHGT, OS, LB, AZ, and MK designed the experiment. YA, MT, and JDB developed the Timepix3 implementation for EELS. NV, YA, SYW, and LHGT performed experiments with support from XL. NV, YA, and LHGT analyzed the data. KW and TT the provided materials. All authors contributed to writing the manuscript.

\section*{Competing interests}
Mathieu Kociak patented and licensed technologies at the basis of the Attolight M\"onch used in this study and is a part time consultant at Attolight. All other authors declare no competing financial interests. 
   
\section*{Supplementary materials}
Materials and Methods\\
Supplementary Text\\
Figs. S1 to S5\\

\newpage
\newpage
	\setcounter{page}{1}
	\renewcommand\thefigure{S\arabic{figure}}
	\renewcommand{\thesection}{S\arabic{section}}
	\setcounter{figure}{0} 
	
\setcounter{section}{0}	

\begin{center}
\section*{Supplementary materials to Cathodolumimnescence excitation spectroscopy: nanoscale imaging of excitation pathways}

\normalsize{Nadezda Varkentina,$^{1\ast}$ Yves Auad,$^{1\ast}$ Steffi Y. Woo,$^{1}$  Alberto Zobelli,$^{1}$\\}
\normalsize{Laura Bocher,$^{1}$ Jean-Denis Blazit,$^{1}$ Xiaoyan Li,$^{1}$ Marcel Tenc\'e,$^{1}$\\}
\normalsize{Kenji Watanabe,$^{2}$ Takashi Taniguchi,$^{3}$ Odile St\'ephan,$^{1}$ Mathieu Kociak,$^{1\dagger}$\\}
\normalsize{Luiz~H.~G.~Tizei,$^{1\dagger}$\\}

\normalsize{$^{1}$Universit\'e Paris-Saclay, CNRS, Laboratoire de Physique des Solides,} \\
\normalsize{91405, Orsay, France}\\
\normalsize{$^{2}$Research Center for Functional Materials, National Institute for Materials Science,} \\
\normalsize{1-1 Namiki, Tsukuba 305-0044, Japan}\\
\normalsize{$^{3}$International Center for Materials Nanoarchitectonics,}\\
\normalsize{National Institute for Materials Science, 1-1 Namiki, Tsukuba 305-0044, Japan}\\
\normalsize{$^\ast$These authors contributed equally}\\
\normalsize{$^\dagger$To whom correspondence should be addressed;}\\
\normalsize{E-mail:  luiz.galvao-tizei@universite-paris-saclay.fr/}\\
\normalsize{mathieu.kociak@universite-paris-saclay.fr}

\end{center}
\setcounter{section}{0}
\section{Methods}

Coincidence EELS-CL experiments were performed on a modified Vacuum Generators (VG) HB501 scanning transmission electron microscope (STEM) equipped with a cold field emission source, an Attolight M\"onch light collection system, a liquid nitrogen cooled sample stage, and a Cheetah Timepix3 (manufactured by Amsterdam Scientific Instruments) event-based direct electron detector. More details about the experimental setup and event based detection can be found in Auad et al. \cite{Auad2022b}. Beam current in coincidence measurements was typically from 1 to 10 pA and convergence half-angle of 7.5 mrad was used. 

High energy resolution EELS and CL measurements and high angle annular dark-field (HAADF) imaging (Fig. \ref{Exc-Dec_Pathways}A-B) were performed on monochromated and Cs-corrected ChromaTEM  modified Hermes200 STEM from NION. Spatially resolved data are acquired by scanning a subnanometer electron beam on the sample. Beam current in the order of 200 pA and convergence half-angle of 25 mrad were used for the experiments. EELS dispersion was set to either 25 meV per channel for low losses (Fig. \ref{Exc-Dec_Pathways}A-B) or 270 meV per channel for core-losses (Fig. \ref{Exc-Dec_Pathways}B). CL used a M\"onch system from Attolight, fitted with a diffraction grating giving an wavelength resolution of 0.34 nm (about 2 meV at 500 nm in wavelength). 

Experiments were performed with 60 and 100 keV electron kinetic energy. \textit{h}-BN flakes suffered damage at 100 keV on experiments on the VG microscope, where the sample chamber vacuum conditions are degraded (higher pressure and water partial pressure) in comparison to the ChromaTEM microscope.

Data acquisition was handled with Nion Swift 1.5 (ChromaTEM) and 1.6 (VG) python-based microscopy-control application.

\section{Samples description}

The first sample consists of gold  silica core-shell nanospheres from nanoComposix. Their total diameter measured by TEM is $140 \pm 10$ nm (100 nm core with a 20 nm shell) as stated by manufacturer. The nanospheres solution were further diluted in spectral quality ethanol in proportion 1:2. One drop of the final solution was then drop cast on a conductive lacey carbon film (tens of nanometers thick) supported on copper TEM grid (agar Scientific) and the extra volume was absorbed by filter paper (Whatman). 

The second sample is a Carbon-13 doped hexagonal Boron Nitride (\textit{h}-BN) \cite{Taniguchi2007}. Thin \textit{h}-BN flakes were prepared by liquid phase exfoliation from \textit{h}-BN monocrystals dispersed in 1 mL of spectral quality isopropanol (Carlos Erba), and then sonicated for 15 minutes. Three drops of the solution were then successively drop cast onto the TEM grids (suspended by tweezers) containing the Au/SiO$_2$ nanospheres. The grid was left to dry until total evaporation of the solvent.

Thin h-BN flakes (below 50 nm) areas were chosen, to ensure that fast electrons suffer statistically at most one scattering event on the sample \cite{E96} (sample relative thickness below one mean free path for inelastic scattering).

\section{Description of coincidence seeking algorithm}

Before applying the coincidence-seeking algorithm, electron events in Timepix3 were cluster-corrected following a procedure that can be found in ref. \cite{Auad2022b}. The time of arrival of each electron event is compared with the time of arrival list from the photon events (after both being sorted by time). A coincident electron is found when its time lies within a given time interval ($\pm25$ ns or $\pm50$ in this work) centered at the first-matching photon time. Instead of pairing every electron with all the elements of the photon list, a sliding window algorithm is performed, which allows to increase performance by as much as a factor of 10 due to the effective photon list's reduced size. The algorithm standard output consists of a list of energy-loss indices along their associated time delay. For the hyperspectral dataset reconstruction (Fig \ref{Spim_hBN}), supplementary events from the microscope scanning unit is used. This procedure can be found in detail in ref. \cite{Auad2022b}. Time delay ($\Delta t$) was set to zero at the maximum of coincidence counts. Its absolute value is meaningless as it includes electronic and propagation delays.

\section{Data processing and availability}
All data was processed using the following Python libraries: Numpy 1.21.5, Matplotlib 3.4.3, Scipy 1.7.3, Hyperspy 1.6.5 \cite{francisco_de_la_pena_2020_4294676}. The raw data processing code is available in Zenodo \cite{Auad2022_6346261}.

All published data are available at the Zenodo repository with the following DOI: 10.5281/zenodo.6803239. An example of raw data for low loss and core loss CLE of h-BN is provided in the same repository.
\newpage
\section{Projections of the 2D histogram of Fig. \ref{CLE_hBN}A}

\begin{figure}[H]
\begin{center}
  \includegraphics[width=5 in]{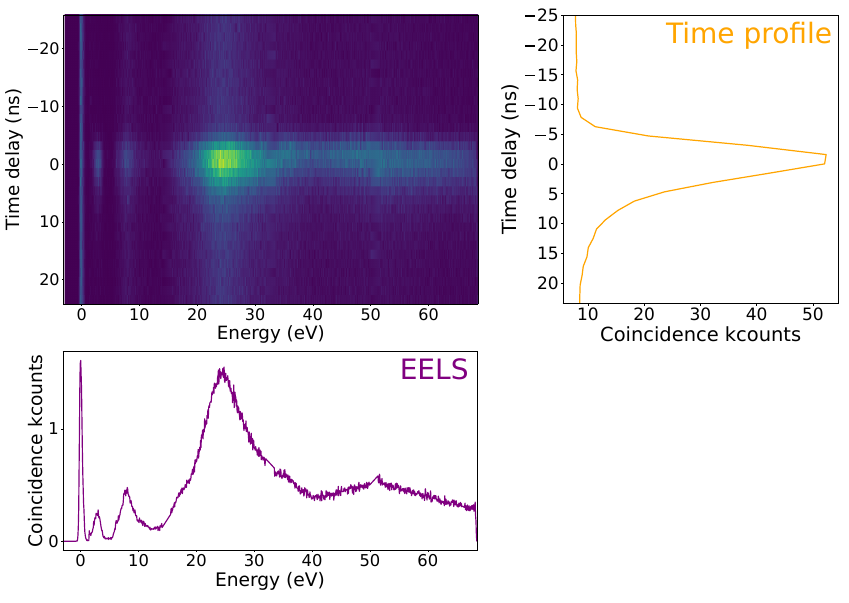}
  \caption{\textbf{Projections along the energy and time delay axes of the 2D histogram of Fig. \ref{CLE_hBN}A:} the 2D time delay/energy histograms, as the one of Fig. \ref{CLE_hBN}A, depict the probability of a specific energy loss to lead to a photon emission after a given time delay. To facilitate the visualization, projections along the energy (right) and the time delay (bottom) axes are shown.}
  \label{Fig2C_Projections}
\end{center}
\end{figure}

\newpage

\section{EELS, CLE, relative QE of nanosphere in Fig. \ref{CLE_Plasmons} and \ref{CLE_rQE}A}

\begin{figure}[H]
\begin{center}
  \includegraphics[width=3 in]{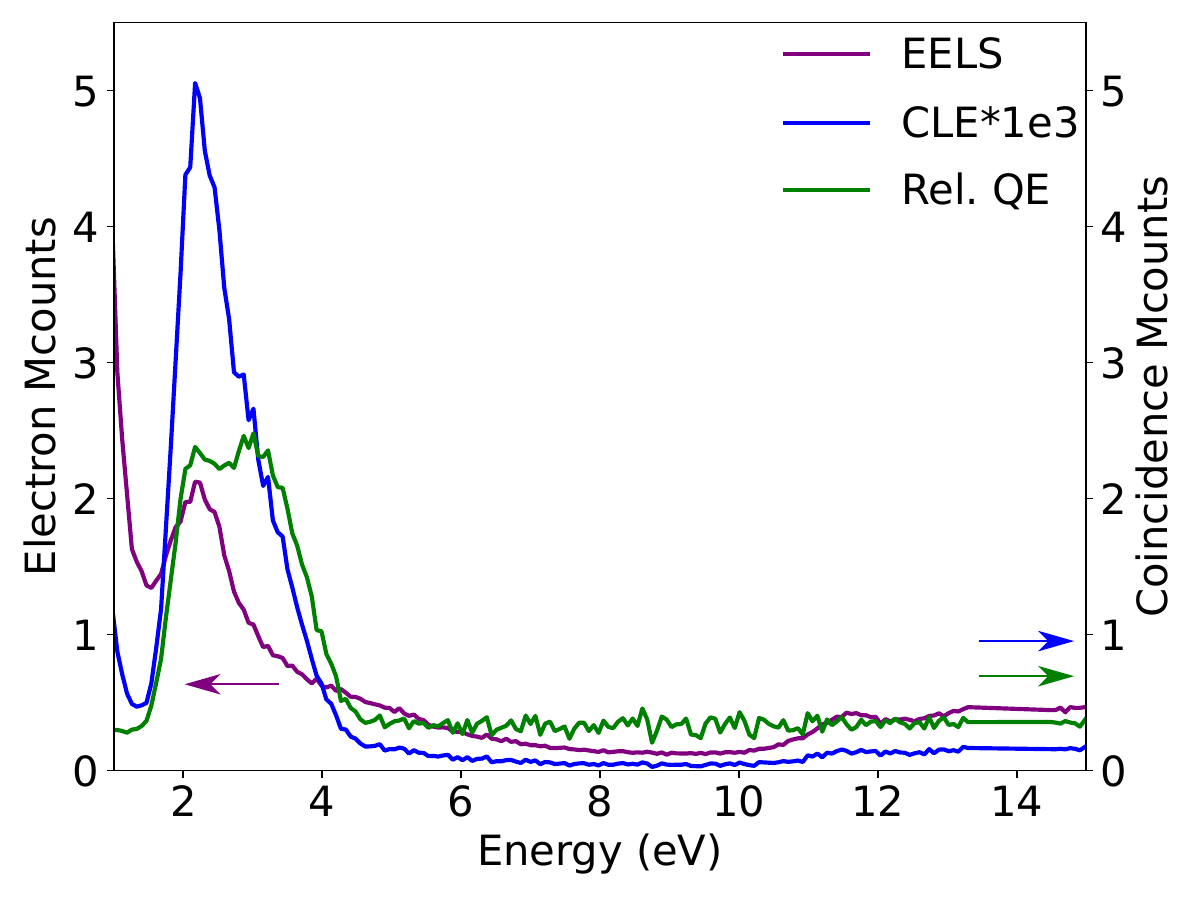}
  \caption{\textbf{EELS, CLE and relative QE of the nanosphere in Fig. \ref{CLE_Plasmons}A and \ref{CLE_rQE}A:} The three spectra were plotted in a narrower energy band and in the same graph to show the double peak structure in the relative quantum efficiency (green). The first peak appears due to the SP resonance, seen in the EELS spectrum, while the second occurs due to TR. Its presence is already indicated in the CLE spectrum (blue), by a shoulder on the SP resonance.}
  \label{SameAxisNanoSFig2}
\end{center}
\end{figure}

A new graph with the same spectra as in Fig. \ref{CLE_Plasmons}B and \ref{CLE_rQE}A  is presented in Fig. \ref{SameAxisNanoSFig2}, to show in detail the double peaks due to SP and TR in the relative quantum efficiency of a Au/SiO$_2$ nanosphere.

\newpage

\section{CLE nanosphere in impact and aloof geometry}
EELS, CLE and relative quantum efficiency for a nanosphere in the aloof (electron beam outside the SiO$_2$ surface) and impact (electron beam on the SiO$_2$ layer) geometries.
\begin{figure}[H]
\begin{center}
  \includegraphics[width=6 in]{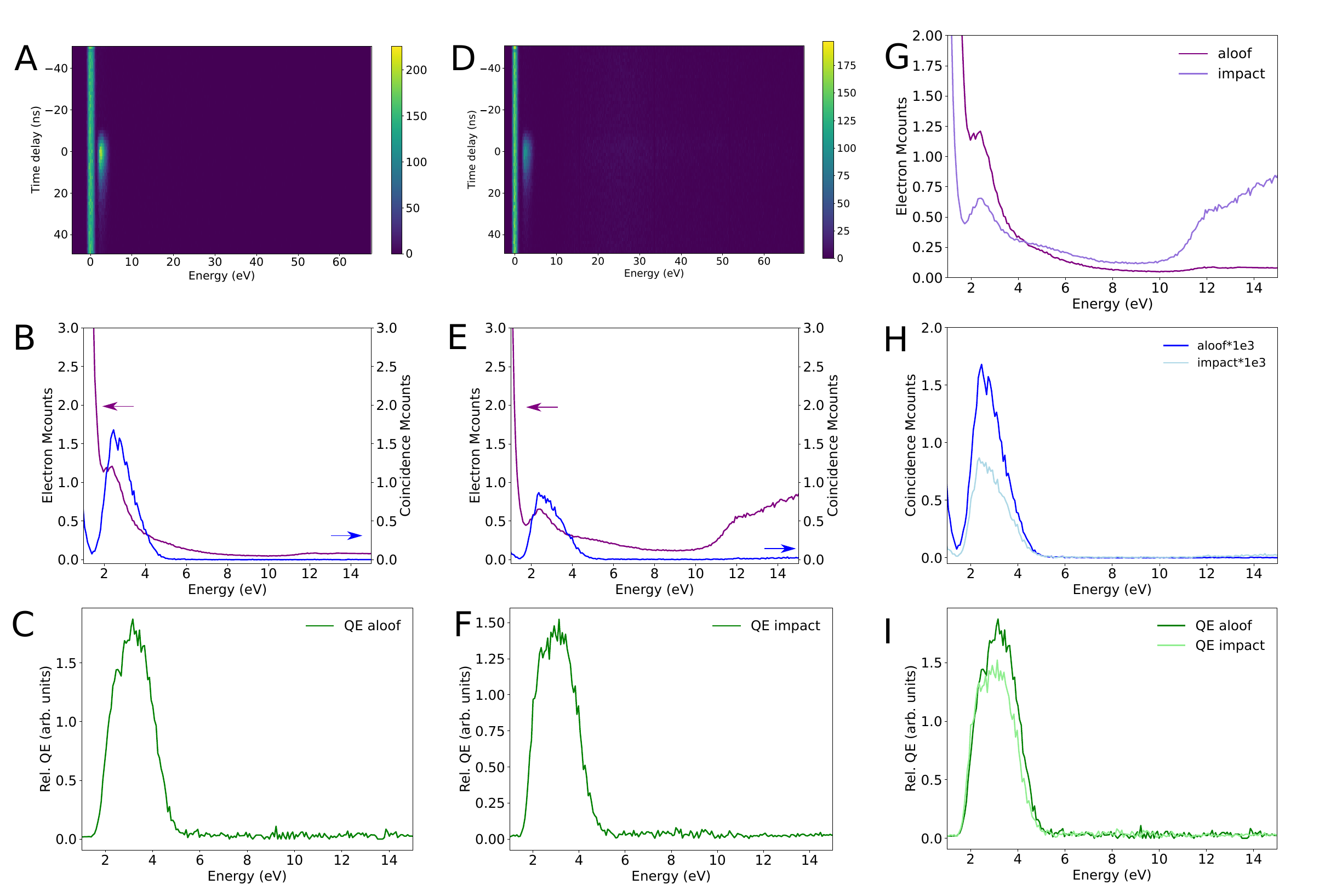}
  \caption{ \textbf{EELS, CLE and relative QE of a nanosphere in aloof and impact geometries:} \textbf{(A-C)} EELS, CLE and relative QE for an Au/SiO$_2$ nanosphere in aloof geometry. \textbf{(D-F)} EELS, CLE and relative QE in impact geometry. \textbf{(G-I)} Comparison of  EELS, CLE and relative QE between aloof and impact geometries.}
  \label{CLE_nanosphere_aloof_impact}
\end{center}
\end{figure}

\newpage

\section{CLE of \textit{h}-BN up to core losses}
\begin{figure}[H]
\begin{center}
  \includegraphics[width=5 in]{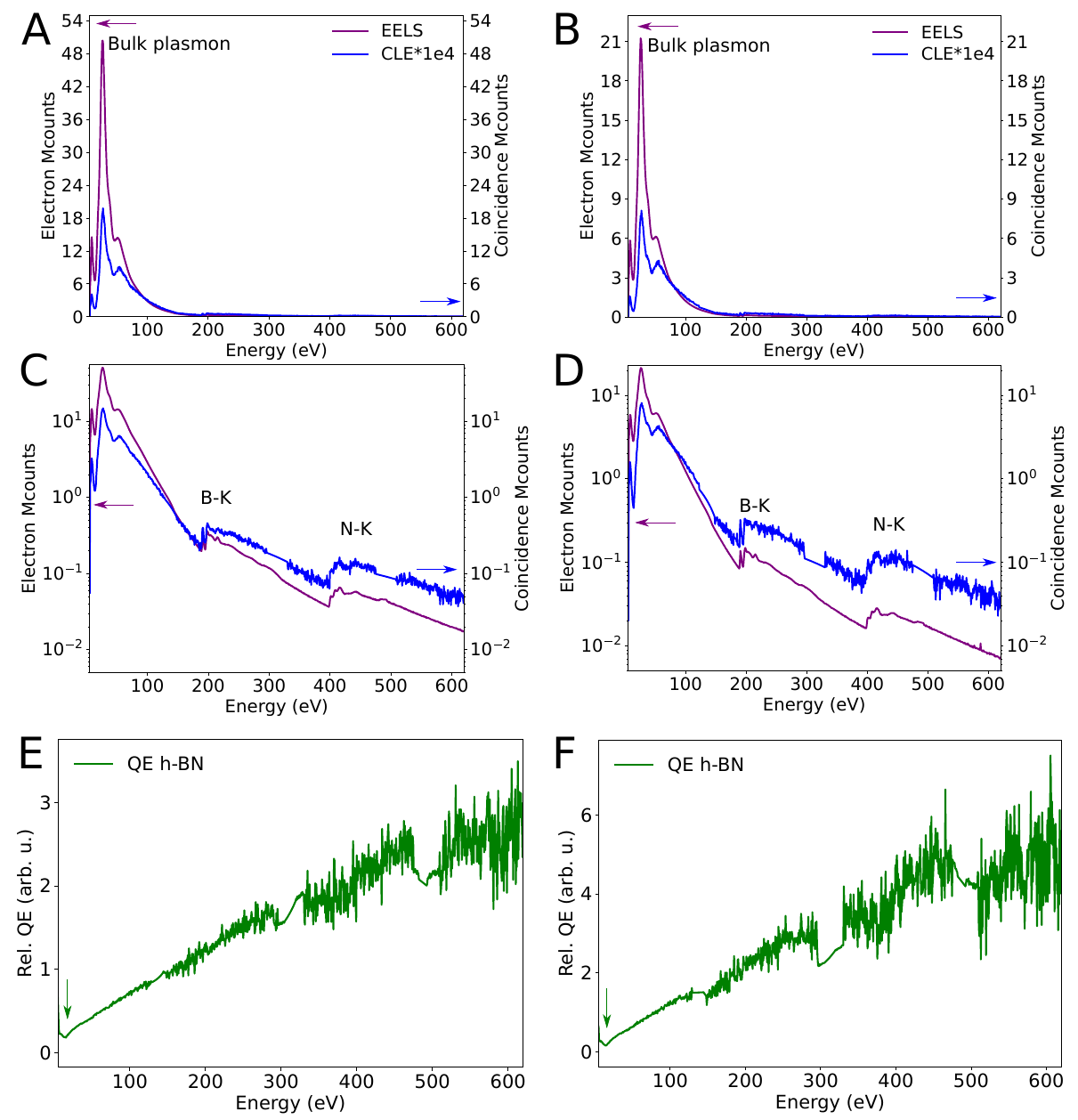}\\
  \caption{\textbf{(A-B)} EELS and CLE of two \textit{h}-BN flakes measured up to 620 eV. \textbf{(C-D)} Same spectra as in A-B but with logarithmic scales. B-K and N-K mark the boron and nitrogen EELS K edges. \textbf{(E-F)} Relative QE for the two measured flakes. The minimum of relative QE is visible at around 15 eV, as well as the decrease in relative QE between 6.5 and 15 eV (pointed by green vertical arrows). CLE integrated time range $\pm 10 $ ns.}
  \label{CLE_core-loss_hBN}
\end{center}
\end{figure}

The relative QE of \textit{h}-BN for the 4.1 eV emission has a linear dependence up to 620 eV shown in Fig. \ref{CLE_core-loss_hBN}, within the currently achievable signal-to-noise ratio. B and N K-edge fine structure is visible in the CLE spectra.  No spectral fine structure appears in the relative QE, but the noise level is still high. The minimum of relative QE is visible at around 15 eV, as well as the decrease in relative QE between 6.5 and 15 eV (pointed by green vertical arrows in \ref{CLE_core-loss_hBN}E-F). Panels A-C are the same as those in Figs. \ref{CLE_hBN}C-D and \ref{CLE_rQE}A for comparison.

\section{CLE of \textit{h}-BN with a 3.65-4.1 eV bandpass filter}

\begin{figure}[H]
\begin{center}
  \includegraphics[width=5 in]{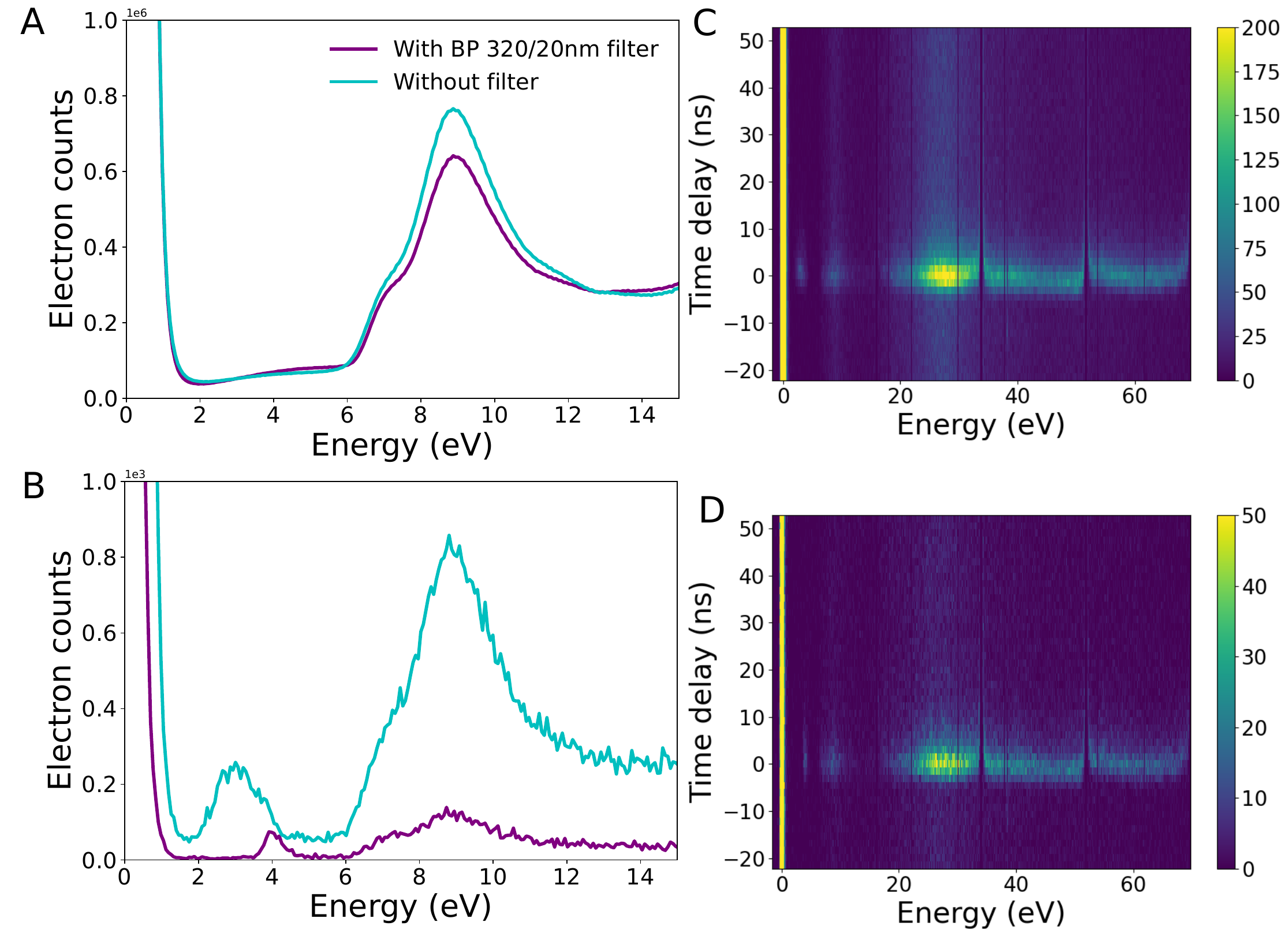}\\
  \caption{\textbf{(A-B)} EELS and CLE of an \textit{h}-BN flake measured without a filter (PMT response in the 2.0 to 5.0 eV range) and with a bandpass filter (3.65 to 4.1 eV). \textbf{(C-D)} 2D histograms for the configurations with and without bandpass filter, respectively.}
  \label{CLE_hBNFiltered}
\end{center}
\end{figure}

The TR photons described in the text show a broad energy spectrum. These are filtered out, if a bandpass is used (\ref{CLE_hBNFiltered} ).

\section{Spurious coincidence events}
\label{Spurious}
First of all, both detectors (a PMT and a Timepix3) used contain noise. For the PMT, these can be remaining photon arriving at the detector from difference sources: ambient light entering the detector, thermal excitation in the photocathode.

Moreover, the electron source used is stochastic, which can lead to spurious correlations. For the current range (10 pA) in our experiments there is 1 electron every 25 ns, on average. Having 2 electrons withing the same detector resolution (our time response function is about 10 ns) would lead to spurious correlations. As a cold FEG is poissonian (for long time delays compared to the emission process) the average number of electron in a 10 ns bin, i.e., during the time between two detection events unresolved by our acquisition chain is:
$\lambda = 10/25$ e/ns (the 10 comes from an estimate of our PSF)
then the probability to detect within the time bin 0 electrons is:
$P(k= 0, \lambda =0.4) = 0.67$
For one electron the probability is:
$P(k= 1, \lambda =0.4) = 0.25$
And that for more than one electron is:
$1-P(k=0) – P(k=1) = 0.08$.
This tells us that about 8 $ \%$ of time bins can contain more than one electron, therefore leading to spurious correlations. That is $0.08/0.033 = 0.22$ of all time bins containing one electron or more. To decrease these undesired events and therefore increase the signal to background in CLE, some options are possible. This could be done using a pulsed gun with repetition rate below 1/10 ns$^{-1}$, using lower emission currents or improving the time response of the detection setup.

\end{document}